\documentclass[
superscriptaddress,
groupedaddress,
reprint,
nofootinbib,
amsmath,amssymb,
aps,
prd,
floatfix,
showkeys,
]{revtex4-2}
\usepackage[utf8]{inputenc}
\usepackage{graphicx}
\usepackage{dcolumn}
\usepackage{bm}
\usepackage{hyperref}
\hypersetup{
    colorlinks=true,
    allcolors=blue,
}

\newcommand{\eq}[1]{Eq.~\eqref{#1}}
\DeclareMathOperator{\arcsinh}{arcsinh}
\DeclareMathOperator{\arccosh}{arccosh}
\def\mybf#1{{\bfseries #1}}

\usepackage[normalem]{ulem}
\newif\ifrevision
\revisionfalse

\ifrevision
  \newcommand{\add}[1]{\textcolor{blue}{#1}}
  \newcommand{\del}[1]{\textcolor{red}{\sout{#1}}}

  \newcommand{\note}[1]{\textcolor{teal}{[Note: #1]}}
\else
  \newcommand{\add}[1]{#1}
  \newcommand{\del}[1]{}
  \newcommand{\note}[1]{}
\fi

\begin{document}

% \title{A Novel Solution to Bumblebee Gravity}
\title{Full Classification of Static Spherical Vacuum Solutions to Bumblebee Gravity with General VEVs}
% \title{Full Classification of Static Spherical Vacuum Solutions in Bumblebee Gravity: Identifying the Exact Schwarzschild Metric}
% \title{Exact Schwarzschild Solution with Non-Vanishing Matter Distributions from Bumblebee Gravity}
\author{Jie Zhu}
 \email{jiezhu@cqu.edu.cn}
 \affiliation{Department of Physics, Chongqing University, Chongqing 401331, P.R. China}

\author{Hao Li}
  \email{Corresponding author: haolee@cqu.edu.cn}
   \affiliation{Department of Physics, Chongqing University, Chongqing 401331, P.R. China}

\date{\today}

\begin{abstract}
    The static spherical vacuum solution in a bumblebee gravity model where the bumblebee field \(B_\mu\) has a two-component space-like, light-like, and time-like vacuum expectation value \(b_\mu\) is studied.
    Based on the results, we present a comprehensive classification of the static spherical vacuum solutions in bumblebee gravity with general vacuum expectation values.
    We find that the model becomes degenerate for a specific set of parameter combinations, where the solution can be characterized by an arbitrary function, 
    which indicates that the non-minimally coupled massless vector tensor theory is ill-defined when $\xi=\kappa/2$.
    % This degeneracy indicates that the vector-tensor theory with non-minimal coupling $\alpha A^\mu A_\mu R +\beta A^\mu A^\nu R_{\mu\nu}$ is degenerate and ill-defined when $\beta = 1/4$ and $\alpha \neq -1/4$.
    We also find that contrary to the situation in general relativity, the bumblebee gravity admits the exact Schwarzschild solution with {\emph{non-zero}} matter distributions of certain forms.
    The implications of this result are discussed, suggesting that the experimental constraints within the solar system would be invalid.
\end{abstract}


\maketitle

\section*{Introduction}

% General Relativity~(GR) and the Standard Model~(SM) of particle physics are the most successful theories describing all four fundamental forces of nature.
% However, there are theoretical tensions between GR and SM, and to reconcile them, several candidates of quantum gravity~(QG) theories have already been proposed.
% Generally, the onset of significant effects of QG is expected to happen at the Planck scale~(\(E_{Pl}\sim 10^{19}~\text{GeV}\)), which is far beyond our reach for current experiments.
% Although direct detection of QG effects seems to be unlikely at present, it is suggested that there exists the possibility that certain kinds of remnant signals of QG could be observed at energy scales much lower than the Planck scale.
% One of such signals is the violation of Lorentz invariance.
General Relativity~(GR) together with the Standard Model~(SM) of particle physics forms the cornerstone of our current understanding of the four fundamental interactions in nature. 
Despite their remarkable success, these two frameworks are conceptually incompatible in certain regimes where effects of quantum mechanics and spacetime curvature cannot be considered separately, motivating the pursuit of a unified theory of quantum gravity (QG). 
The characteristic energy scale at which QG effects are expected to become significant is the Planck scale~(\(E_{\text{Pl}}\sim 10^{19}~\text{GeV}\)), which lies far beyond the reach of present-day experiments. 
While a direct observation of QG phenomena remains practically impossible, it has been argued that indirect imprints of QG might persist at energies much lower than the Planck scale. 
Among the potential low-energy manifestations, a particularly intriguing one is the possible violation of Lorentz invariance.

% In recent decades, increasing interest has been directed toward possible violations of Lorentz symmetry, driven by attempts to formulate a consistent theory of quantum gravity and to understand potential deviations from GR at high energies. Among various approaches, effective field theories incorporating spontaneous Lorentz violation have proven especially fruitful. In this context, Bumblebee gravity has emerged as a minimal yet nontrivial extension of GR, wherein a vector field acquires a nonzero vacuum expectation value (VEV), leading to spontaneous breaking of local Lorentz invariance and diffeomorphism invariance in a controlled manner~\cite{Kostelecky:1988zi,Kostelecky:2003fs,Bluhm:2004ep,Bailey:2006fd,Maluf:2013nva}.
In recent years, growing attention has been devoted to possible violations of Lorentz symmetry, motivated by efforts to construct a consistent framework for quantum gravity and to probe potential deviations from General Relativity (GR) at high-energy scales. 
Among the various theoretical approaches, effective field theories that incorporate spontaneous Lorentz violation have proven to be particularly insightful. 
Within this framework, the so-called \emph{bumblebee gravity} serves as a simple yet nontrivial extension of GR, in which a vector field develops a non-vanishing vacuum expectation value~(VEV), thereby inducing spontaneous breaking of local Lorentz and diffeomorphism invariance in a controlled and consistent fashion~\cite{Kostelecky:1988zi,Kostelecky:2003fs,Bluhm:2004ep,Bailey:2006fd,Maluf:2013nva}.

The Bumblebee model typically introduces a vector field $B_{\mu}$, non-minimally coupled to gravity and
governed by a potential $V(B_\mu B^\mu\pm b^2)$, which determines the vacua of the theory at the classical level.
The vacua, determined by the vacuum expectation \(\langle B^\mu\rangle\) such that \(V|_{B_\mu=\langle B_\mu\rangle}\) defines the minima of the potential, do not transform as scalars, thus signaling the spontaneous breaking of Lorentz symmetry.
% enforces the constraint $B_\mu B^\mu=\pm b^2$ at the level of field equations.
This spontaneous Lorentz symmetry breaking results in modifications to the Einstein field
equations, leading to potentially observable signatures in gravitational phenomena. 
\add{Various aspects of the bumblebee model have been investigated in recent years, including black-hole physics~\cite{Liu:2024axg,  Xu:2023xqh, Mai:2023ggs, Chen:2020qyp, Liu:2022dcn, Zhang:2023wwk, Deng:2025uvp, Ding:2020kfr}, cosmology~\cite{Lai:2025nyo, Xu:2025heq}, and gravitational waves~\cite{Liang:2022hxd}. }
In previous studies, an exact Schwarzschild-like solution in bumblebee gravity is proposed~\cite{Casana:2017jkc}.
Also, an exact Kerr-like solution is found~\cite{Ding:2019mal}.
In these solutions, the configurations of the bumblebee field only admit space-like VEVs, and the temporal component of the bumblebee field is fixed $\langle B_t\rangle =0$~\footnote{We are interested in spherically symmetric solutions, so the angular components of \(\langle B_\mu\rangle\) are ignored: \(\langle B_\theta\rangle=\langle B_\phi\rangle=0\) henceforth.}.
Similar results are also obtained and analyzed in detail within metric-affine bumblebee models~\cite{Filho:2022yrk,AraujoFilho:2024ykw,AraujoFilho:2025hkm}.
In our previous work, we explored static spherical vacuum solutions of the Bumblebee gravity model with a time-like vacuum expectation~\cite{Li:2025bzo}, and the radial component of the field configuration is neglected, $\langle B_r\rangle=0$.
% But for both of the solutions, the configurations of the bumblebee field only admit space-like VEVs, i.e., $\langle B^\mu B_\mu\rangle > 0$.
In this work, we systematically explore the general static spherical vacuum solutions of the Bumblebee gravity model with general types of spherically symmetric VEVs.
We show the details of the systematic solving procedure.
The solution procedure demonstrates that we have exhaustively identified all possible solutions across the entire parameter space, based on which we propose a classification of the solutions.
With the complete classification, we identify the {\emph{exact}} Schwarzschild solution accompanied by {\emph{non-trivial}} matter distributions.
The important implications of this result are analyzed in detail.

% This literature is organized as follows.
% In Sec.~\ref {sec:2}, we briefly introduce the action and the equation of motion for bumblebee gravity.
% In Sec.~\ref {sec:3}, we try to solve the equations of motion in a static spherical field configuration with the bumblebee field obtaining time-like VEVs, and we find that for general VEVs of this type there is no solution for the equations, unless $b=\sqrt{2/\kappa}$.
% In Sec.~\ref{sec:4}, we solve the equations when $b=\sqrt{2/\kappa}$, and find two kinds of non-trivial solutions.
% Sec.~\ref{sec:5} is the discussion and the summary.
% In this literature, we will adopt the metric signature $(- + + +)$ and also all quantities involved are expressed in natural units $(\hbar=c=1)$.

% \section{Brief Introduction of Bumblebee gravity}\label{sec:2}

\section*{The Bumblebee Model}

The action of Bumblebee gravity can be expressed as~\cite{Casana:2017jkc}~\footnote{We have omitted additional terms proportional to the cosmological constant \(\Lambda\), \(B^\mu B_\mu R\), \(\nabla_\mu B_\nu\nabla^\mu B^\nu\) and \((\nabla_\mu B^\mu)^2\)}
\begin{align}
    S=&\int d^4x\sqrt{-g}\left(\frac{1}{2\kappa}R+\frac{\xi}{2\kappa}B^\mu B^\nu R_{\mu\nu}-\frac{1}{4}B^{\mu\nu}B_{\mu\nu}-V\right)\nonumber\\ &+S_\mathrm{m},\label{action}
\end{align}
where $g$ is the determinant of the metric $g_{\mu\nu}$, $\kappa\equiv8\pi G$ with $G$ being the gravitational constant, $S_{\mathrm{m}}$ represents the action for matter fields~(except the bumblebee field), which is of no interest in this work, $B_\mu$ is the bumblebee field, and the field strength tensor is as usual $B_{\mu\nu}=\partial_{\mu}B_{\nu}-\partial_{\nu}B_{\mu}$.
In bumblebee theories, the potential $V$ is selected to provide a non-vanishing VEV for $B_\mu$, and could have the following general functional form
\begin{equation}
    V\equiv V(B^\mu B_\mu + s b^2),\label{potential}
\end{equation}
where $b$ is a positive real constant, and $s=\pm 1$ or \(s=0\), determining whether the expectation value of $B_\mu$ is time-like, space-like or light-like respectively. 
Correspondingly, in the literature it is usually assumed that \(V\) has (at least one of) its minimum/maximum at \(0\), thus %~\footnote{The condition for \(V''(0)\) plays no role in this work.}
\begin{equation}
    V(0)=0,\ \text{and}\ V'(0)=0.\label{vacuumcondition}
\end{equation}
Obviously, the VEV of the bumblebee field is determined when $V(B^\mu B_\mu + s b^2)=0$, implying that
\begin{equation}
    B^\mu B_\mu+s b^2=0, 
\end{equation}
The above equation indicates a generally non-zero vacuum expectation value
\begin{equation}
    \langle B^\mu\rangle=b^\mu,
\end{equation}
so that $b_\mu b^\mu + s b^2=0$, and
\begin{equation}
    \begin{aligned}
     \left.V\right|_{B_\mu=b_\mu}&=0,\\
    \left.\frac{dV}{d(B^\lambda B_\lambda)}\right|_{B_\mu=b_\mu}&=0.
    \end{aligned}
\end{equation}

We are interested in the vacuum solutions, namely $(T_m)_{\mu\nu}=0$, in the sense that no other kind of matters exists, besides the bumblebee field which might be composed of some type of unknown matters. 
Provided with the above condition for $V$ and $B_\mu$ replaced exactly by its VEV $b_\mu$, {\it i.e.,} the degrees of freedom of the bumblebee field will not be excited, the equation of motions for $g_{\mu\nu}$ is
\begin{equation}
    R_{\mu\nu}-\frac{1}{2}R\,g_{\mu\nu}-\kappa T^B_{\mu\nu}=0,\label{eq:einsteineq}
\end{equation}
where 
\begin{equation}
    \begin{aligned}T_{\mu\nu}^{B}=&-B_{\mu\alpha}B_{\nu}^{\alpha}-\frac{1}{4}B_{\alpha\beta}B^{\alpha\beta}g_{\mu\nu}
    % -Vg_{\mu\nu}+2V^{\prime}B_{\mu}B_{\nu}
    \\&+\frac{\xi}{\kappa}\left[\frac{1}{2}B^{\alpha}B^{\beta}R_{\alpha\beta}g_{\mu\nu}-B_{\mu}B^{\alpha}R_{\alpha\nu}-B_{\nu}B^{\alpha}R_{\alpha\mu}\right]\\&+\frac{1}{2}\nabla_{\alpha}\nabla_{\mu}\left(B^{\alpha}B_{\nu}\right)+\frac{1}{2}\nabla_{\alpha}\nabla_{\nu}\left(B^{\alpha}B_{\mu}\right)\\&-\frac{1}{2}\nabla^2\left(B_\mu B_\nu\right)-\frac{1}{2}g_{\mu\nu}\nabla_\alpha\nabla_\beta\left(B^\alpha B^\beta\right).
    \end{aligned}
\end{equation}
Taking the trace of \eq{eq:einsteineq}, we get $R=-\kappa T^B$, and thus the above equation can be expressed as
\begin{equation}
    R_{\mu\nu}-\kappa(T^B_{\mu\nu}-\frac{1}{2}g_{\mu\nu}T^B)=0. \label{eq:gt}
\end{equation}
On the other hand, the equation of motion for \(B_\mu\) is
\begin{equation}
    % \frac{\xi}{\kappa}B^\mu R_{\mu \nu}+\nabla_\mu\nabla^\mu B_\nu-\nabla_\mu\nabla_\nu B^\mu=0. 
    \nabla^{\mu}B_{\mu\nu}+\frac{\xi}{\kappa}B^\mu R_{\mu \nu}=0.\label{eq:b}
\end{equation}

The general static spherical background $b_\mu$ can be expressed as
\begin{equation}
    b_\mu=(b_t,b_r,0,0),\label{eq:b_s}
\end{equation}
where $b_t$ and $b_r$ are functions of $r$.
Without loss of generality, we assume the static spherically symmetric metric to be~\footnote{Since we have included an unknown function \(R(r)\) in front of \(d\Omega^2\), the factor of \(dr^2\) can always be chosen to be \(e^{-2\alpha(r)}\) by a suitable redefinition of \(r\).}
\begin{equation}
    ds^2=-e^{2\alpha(r)}dt^2+e^{-2\alpha(r)}dr^2+R(r)^2d\Omega^2. \label{eq:met}
\end{equation}
Substituting \eq{eq:met} and \eq{eq:b_s} into \eq{eq:gt} and \eq{eq:b}, together with $b_\mu b^\mu = s b^2$, we can solve and obtain the general static spherical solutions.
For clarity, the details of the systematic solving procedure can be found in the appendix, and in the following we summarize the results. For simplicity, we define $\ell\equiv\xi b^2$.

\section*{Results}

\begin{description}
\item[Case I] The parameters $\xi$ and $b$ are general. 

In this case, the solution for the metric is
\begin{equation}
    ds^2= -\frac{1}{1+\alpha \ell}fdt^2+(1+\alpha \ell)f^{-1}dr^2+r^2d\Omega^2,
\end{equation}
where $f(r)=1-R_s/r$, and $\alpha$ and $R_s$ are constants of integration.
It is noteworthy that this form of solution is indeed universal in the sense that no matter whether the background \(b_\mu\) is space-like, light-like, or even time-like.
% The space-like, light-like, and time-like VEVs for $b_\mu$ all exist in this case. 
% In this case, no matter whether \(b_\mu\) is space-like, light-like, or time-like, a corresponding solution always exists.
% To be more inspiring, we first introduce another constant \(\beta\), which will be related to \(\alpha\) differently for different solutions.
%
Correspondingly, we could exactly solve the VEVs, and to be more clear, we first introduce another constant \(\beta\) which will be related to \(\alpha\).
Then for space-like VEVs, with $\alpha=\beta^2+1$, the solution is
\begin{equation}
\begin{aligned}
    b_t=&  b\frac{\beta}{\sqrt{1+\alpha \ell}},\\
    b_r=& b\sqrt{1+\alpha \ell} \frac{\sqrt{\beta^2+f}}{f}.
\end{aligned}
\end{equation}
    % where $\alpha=\beta^2+1$.
%    
For light-like VEVs, with $\alpha=\beta^2$, the solution is
\begin{equation}
\begin{aligned}
    b_t=& b \frac{\beta}{\sqrt{1+\alpha \ell}},\\
    b_r=& b \frac{\beta \sqrt{1+\alpha \ell}}{f}.
\end{aligned}
\end{equation}
    % where $\alpha=\beta^2$.
%    
For time-like VEVs, with $\alpha=\beta^2-1$, the solution is
\begin{equation}
\begin{aligned}
b_t=&  \frac{b}{\sqrt{1+\alpha \ell}}\beta,\\
b_r=& b\sqrt{1+\alpha \ell} \frac{\sqrt{\beta^2-f}}{f}.
\end{aligned}
\end{equation}
    % where $\alpha=\beta^2-1$.
    
In all of the solutions, $b_t$ is a constant, and $b_r$ cannot be identically zero in general.
The only exception is for light-like VEVs and \(\beta=0\).
It is interesting that the type of VEVs can be read off from the constant \(\alpha\) represented by \(\beta\): for space-like VEVs, \(\alpha=\beta^2+1\) is always positive; for light-like VEVs, \(\alpha=\beta^2\) is also positive with non-trivial $b_\mu$; while for time-like VEVs \(\alpha=\beta^2-1\), of course, may be negative.
Also remarkable is that,
in the space-like case, if $\beta=0$, then $\alpha=1$, and the solution reduces to the solution obtained in Ref.~\cite{Casana:2017jkc}.
Besides,
it is worth noting that when $\beta= 1$ in the time-like case, $\alpha=0$, and the solution of the metric reduces to the Schwarzschild black hole as well.  
However, the field configuration $b_\mu$ is \(b_\mu = b(1, \frac{\sqrt{R_s r}}{r-R_s},0,0)\), which is non-trivial and even not zero when \(r\to\infty\).
    
\item[Case II] The parameters satisfy $\xi=\kappa/2$.
 
In this case, the solution for the metric is again
\begin{equation}
    ds^2= -\frac{1}{1+\alpha \ell}fdt^2+(1+\alpha \ell)f^{-1}dr^2+r^2d\Omega^2,
\end{equation}
where $f=1-R_s/r$, and $\alpha$ and $R_s$ are constants.
% The difference is that there exist non-trivial solutions for $b_\mu$.
The difference from \mybf{Case I} is that, in this case \(b_t\) can be non-constant, and brings about more degrees of freedom to choose parameters.

Similarly, we introduce two extra parameters \(\beta\) and \(\gamma\), which shall be used to replace \(\alpha\).
For space-like VEVs, with $\alpha=(\beta+\gamma)^2+1$, the solution is
\begin{equation}
\begin{aligned}
    b_t=&  \frac{b}{\sqrt{1+\alpha \ell}}\left(\beta+\gamma f\right),\\
    b_r=& b\sqrt{1+\alpha \ell} \frac{\sqrt{(\beta+\gamma f)^2+f}}{f}.
\end{aligned}\label{eq:s2}
\end{equation}
    % where $\alpha=(\beta+\gamma)^2+1$.
    %
For light-like VEVs, with $\alpha=(\beta+\gamma)^2$, the solution is
\begin{equation}
\begin{aligned}
    b_t=& b \frac{1}{\sqrt{1+\alpha \ell}}\left(\beta+\gamma f\right),\\
    % b_r=& \frac{b_t}{f},
    b_r=& b\sqrt{1+\alpha \ell}\left(\frac{\beta}{f}+\gamma \right).
\end{aligned}\label{eq:l2}
\end{equation}
    % where $\alpha=(\beta+\gamma)^2$.
    %
For time-like VEVs, with $\alpha=(\beta+\gamma)^2-1$, the solution is
\begin{equation}
\begin{aligned}
    b_t=&  \frac{b}{\sqrt{1+\alpha \ell}}\left(\beta+\gamma f\right),\\
    b_r=& b\sqrt{1+\alpha \ell} \frac{\sqrt{(\beta+\gamma f)^2-f}}{f}.
\end{aligned}\label{eq:t2}
\end{equation}
    % where $\alpha=(\beta+\gamma)^2-1$.
Once again whether \(\alpha\) could be negative, zero, or only positive is a signature of the type of VEVs as is discussed in the previous case.
The difference and relationship between {\bfseries Case I} and {\bfseries Case II} now is very clear: they can be interchanged by performing a \emph{duality} transformation \(\beta\leftrightarrow \beta+\gamma f\).
In all of the cases, $b_r$ cannot be identically zero, but it can be constant.
It is noteworthy that in this case $b_{\mu\nu}$ has non-vanishing components 
    % $b_{tr}=-b_{rt}\propto 1/r^2$.
$ b_{tr}=-b_{rt}=-\frac{\gamma  b R_s}{\sqrt{1+\alpha\ell}}\frac{1}{r^2}$.
In analogy, the component \(b_{tr}\) corresponds to an \emph{electric} field with \emph{electric} charge \(Q\propto \gamma b R_s/\sqrt{1+\alpha\ell}\).
Moreover, in the light-like case, if \(\beta+\gamma=0\) the metric reduces to the Schwarzschild one,
while for the time-like case, it is the same when $\beta+\gamma=\pm 1$.
    
\item[Case III] The parameters satisfy $b^2=2/\kappa$.
 In this case, there are exotic solutions when $B_\mu$ admits a time-like VEV, and they are explored in our previous work~\cite{Li:2025bzo}.
The bifurcation of this solution occurs at $b_r\equiv 0$, which can only occur in the time-like case (see Appendix~\ref{sec:timelike}). 
There are two families of solutions, which we summarize as in the following.
\begin{itemize}
\item The first one is
\begin{align}
    ds^2 =& -\left(\frac{r}{R_s}\right)^{\frac{2(2-\ell)}{\ell}}dt^2+\frac{4}{\ell^2}dr^2+r^2 d\Omega^2,\\
    b_t = &\sqrt{\frac{2}{\kappa}}\left(\frac{r}{R_s}\right)^{2/\ell-1},
\end{align}
where $R_s$ is a constant.
\item The second one is
\begin{align}
ds^2 =& -\left(1-\frac{R_s}{\rho}\right)^{2-\ell} dt^2+\left(1-\frac{R_s}{\rho}\right)^{-(2-\ell)}d\rho^2 \nonumber\\
&+\left(1-\frac{R_s}{\rho}\right)^{\ell}\rho^2 d\Omega^2,\\
b_t = &\sqrt{\frac{2}{\kappa}}\left(1-\frac{R_s}{\rho}\right)^{1-\ell/2},
\end{align}
where $R_s$ is a constant.
\end{itemize}
The properties of these solutions have already been thoroughly analyzed in Ref.~\cite{Li:2025bzo}.

\item[Case IV] The parameters satisfy $ b^2=1/\xi$. 
% This means that $B_\mu$ admits a space-like VEV $b=\sqrt{-1/\xi}$ with $\xi<0$ and $\ell=-1$, or $B_\mu$ admits a time-like VEV $b=\sqrt{1/\xi}$ with $\xi>0$ and $\ell=1$.

In this case, when $B_\mu$ admits a time-like VEV, 
the solving procedure in {\bfseries Cases I} and {\bfseries Case II} fails since $\ell=1$, and explicit solutions seem to be unobtainable.
Even though the {\bfseries Case I} solution fulfills the present equations, we may have overlooked certain solution branches.
Assuming that the equations admit a black hole solution with a horizon at $r = R_s$, we can obtain series expansions for the metric and $b_\mu$ by performing a series expansion near the horizon. 
The detail is shown in Appendix~\ref{sec:series}.
The result implies that there is a three-parameter family of Schwarzschild-like solutions, determined by the value of $b_t(R_s)$, $b_t'(R_s)$, and $R_s$.
Since the {\bfseries Case I} solution is a two-parameter family,
solutions beyond {\bfseries Case I} indeed exist under these conditions.
It is easy to check that when $b_t'(R_s)=0$, the solution reduces to the solution in {\bfseries Case I}.
    
\item[Case V] The parameters satisfy $ b^2=-1/\xi$.

Similar to {\bfseries Case IV}, when $B_\mu$ admits a space-like VEV, the series solution suggests a three-parameter family of Schwarzschild-like solutions, determined by the value of $b_t(R_s)$, $b_t'(R_s)$, and $R_s$.
However, different from the time-like case, 
% when $b_t'(R_s)=0$ the solution
the limit $b_t'(R_s) \to 0$ does not yield a physically meaningful solution.

\item[Case VI] The parameters satisfy $\xi=\kappa/2$ and $b^2=2/\kappa$.

The parameters in this case simultaneously satisfy the conditions for {\bfseries Case II}, {\bfseries Case III}, and {\bfseries Case IV}.
In this case, when $B_\mu$ admits a time-like VEV, the solution is parameterized by a function, i.e., for \emph{any function} $G(r)$, we can construct a solution as (see Appendix~\ref{sec:any})
\begin{equation}
\begin{aligned}
    ds^2=&-G(r)dt^2+\frac{1}{H(r)}dr^2+r^2d\Omega^2,\\
    b_t=&b\sqrt{G(r)}V(r),\\
    b_r=&b\frac{\sqrt{V(r)^2-1}}{\sqrt{H(r)}},\\
\end{aligned}
\end{equation}
where
\begin{equation}
\begin{aligned}
H(r)=&c_1 \exp\left(
\int \frac{r \left(G'^2-2 G G''\right)}{G \left(r G'+4 G\right)}dr\right),\\
V(r)=&\frac{\pm 1}{r\sqrt{G(r)}}\left(c_2+\int \sqrt{\frac{G(r)}{H(r)}}dr\right),
\end{aligned}
\end{equation}
for some constants \(c_1\) and \(c_2\).
But fixing \(G(r)\) is not necessary, actually we can choose exactly one of \(G(r)\), \(H(r)\), \(b_t(t)\) or \(b_r(r)\) freely, and the others follow by solving the differential equations.
The relationship between $G(r)$ and $H(r)$ results from the condition $R_{rr} = 0$ of the Ricci tensor.
This implies that any static spherically symmetric metric satisfying $R_{rr} = 0$ can be a solution of this system.
Mathematically, this is a manifestation of the degeneration of the system of differential equations derived from the equations of motion.
Specifically, we have more degrees of freedom to choose functions of \(g_{tt}\), \(g_{rr}\), \(b_t\) and \(b_r\) than the number of \emph{independent} differential equations, so we are free to fix one function for \(G(r)\), \(H(r)\), \(b_t(r)\) or \(b_r(r)\), and still are able to solve the others.
For this specific set of model parameters, the vacuum field equations exhibit an physically uncountable degeneracy in their solution space.
Given a fixed VEV \(b_\mu b^\mu\), we \emph{cannot} obtain solutions without ambiguity, even when we require asymptotic flatness of the metric.
However, \(g_{tt}\) and \(g_{rr}\) have real physical effects that can be observed, so ambiguity in them is unacceptable.
Such a scenario implies that the dynamical equations are degenerate and lose their ability to constrain the fields.
It seems that this specific set of parameters should be excluded from the physically viable parameter space, representing a singular boundary where the underlying theory lacks a well-defined dynamical structure.

\end{description}

\section*{Discussions}

As $B_\mu$ is a real-valued vector field, the solution is meaningful when the components of $b_\mu$ are real-valued.
The newly obtained solutions in {\bfseries Case I} and {\bfseries Case II} have horizons at $r=R_s$.
Since $0<f<1$ outside the horizon and $-\infty<f<0$ inside the horizon, in \mybf{Case I}, the space-like solution is not always meaningful inside the horizon, the light-like solution is always meaningful, and the time-like solution is always meaningful when $\beta^2\geq1$.
In \mybf{Case II}, we need $\beta\gamma\leq -1/4$ to make the space-like solution meaningful inside the horizon, and we need $(\beta +\gamma )^2\geq 1$, together with $\beta  \gamma \geq1/4$ or $\gamma(\beta  +\gamma)\leq 1/2$ to make the time-like solution meaningful.
Of course, even if a solution is meaningless on some spacetime patch, like inside the horizon, it might be still acceptable once we can smoothly \emph{glue} it and another solution meaningful on this patch together.

For a specific bumblebee gravity model, the input parameters are the constants in the Lagrangian, i.e., $\xi$ and $sb^2$.
The solutions of the metric are characterized by the combination $\ell=\xi b^2$.
However, for the solutions in \mybf{Case I} and \mybf{Case II}, one more parameter $\alpha$ appears~\footnote{Strictly speaking, when the solutions for \(b_t\) and \(b_r\) are considered, \(\alpha\) should be expressed by more parameters.}, manifesting a characteristic similar to a charge.
In fact, at spatial infinity, the metric does not reduce to the Minkowski form.
Instead, it approaches 
\begin{equation}
    ds^2\simeq-\frac{1}{1+\alpha\ell}dt^2+(1+\alpha\ell)dr^2+r^2d\Omega^2.
\end{equation}
With a redefinition of $t$ and $r$, the metric is
\begin{equation}
    ds^2\simeq-dt^{\prime2}+dr^{\prime2}+\frac{r^{\prime2}}{1+\alpha\ell}d\Omega^2,
\end{equation}
indicating that the spacetime is not asymptotically flat but asymptotically conical.
The constant re-scaling between temporal and spatial components implies a solid-angle deficit and signals that the vacuum geometry itself is non-Minkowskian, arising from a Lorentz-violating background field.
For a specific bumblebee gravity model with fixed $\xi$ and $sb^2$, the parameter $\alpha$ is a charge characterizing the degree of asymptotic compression of the spacetime geometry.
In \mybf{Case II}, for a specific bumblebee gravity model with fixed $\xi$ and $sb^2$ and with the same $\alpha$, there are different field configurations of $b_\mu$ indicating one more degree of freedom, shown as %Eqs.~(\ref{eq:s2}, \ref{eq:l2}, \ref{eq:t2}).
\eq{eq:s2}, \eq{eq:l2} and \eq{eq:t2}.
In these cases, the component $b_{tr}=-b_{rt}\propto \gamma/r^2$, is not zero, suggesting that $\gamma$ is the \emph{electric} charge characterizing the field strength $b_{\mu\nu}$.

% One of the interesting parts of the solutions is the solution parameterized by a function in Case VI.
Note that the equations of motion derived from \eq{action} are identical to those of a massless vector field non-minimally coupled to gravity with Lagrangian
\begin{equation}
    % S=\int d^4x\sqrt{-g}\left(\frac{1}{2\kappa}R+\frac{\xi}{2\kappa}B^\mu B^\nu R_{\mu\nu}-\frac{1}{4}B^{\mu\nu}B_{\mu\nu}\right),
    \mathcal{L}=\frac{1}{2\kappa}R+\frac{\xi}{2\kappa}A^\mu A^\nu R_{\mu\nu}-\frac{1}{4}F^{\mu\nu}F_{\mu\nu},
    \label{eq:vt}
\end{equation}
and consequently, the solutions of the bumblebee gravity model are also solutions to it.
However, in the theory of a massless vector field non-minimally coupled to gravity, the only parameter is the coupling constant $\xi$.
When considering the Bumblebee solutions as solutions to this vector-tensor gravity theory, the parameter $b$ can be treated as a free parameter.
That is, the exotic solutions in \mybf{Case III}, \mybf{Case IV} (when $\xi>0$), and \mybf{Case V} (when $\xi<0$) are special solutions to this vector-tensor gravity theory for a general coupling constant $\xi$.
The solution parameterized by a function as in \mybf{Case VI} indicates that, in this vector-tensor theory, when $\xi=\kappa/2$, the theory is degenerate and ill-defined. 

% This degeneracy can be generalized to the following vector-tensor non-minimally coupled model
% \begin{equation}
%     \mathcal{L}=\frac{1}{2\kappa}R+\alpha A^\mu A_\mu R +\beta A^\mu A^\nu R_{\mu\nu}-\frac{1}{4}F^{\mu\nu}F_{\mu\nu}.\label{eq:gvt}
% \end{equation}
% We can solve the model with an additional constraint $A^\mu A_\mu=-a^2$, and the above Lagrangian reduces to
% \begin{equation}
%     \mathcal{L}=\frac{1}{2\kappa'}R+\beta A^\mu A^\nu R_{\mu\nu}-\frac{1}{4}F^{\mu\nu}F_{\mu\nu},
% \end{equation}
% where $1/\kappa'=1/\kappa-2\alpha a^2$.
% If $\beta=1/4$, the model is degenerate when $a^2=2/\kappa'$, i.e., $a^2=\frac{2}{\kappa(4\alpha+1)}$.
% So when $\beta=1/4$ and $\alpha\neq -1/4$, the original model \eq{eq:gvt} is degenerate and ill-defined, and any static spherically symmetric metric satisfying $R_{rr} = 0$ can be a solution of this model.
% The condition $\beta = 1/4$ was found in previous research, but the crucial implication was disregarded.
% In Ref.~\cite{Chagoya:2016aar}, Chagoya et al. studied a vector-tensor theory featuring a non-minimal coupling between the vector field $A^\mu$ and the Einstein tensor $G_{\mu\nu}$, specifically through the term $\beta A^\mu A^\nu G_{\mu\nu}$. 
% They found that this system admits asymptotically flat solutions only when $\beta = 1/4$.
% Now we know that when $\beta = 1/4$, this system is degenerate and of course admits asymptotically flat solutions.
% Nevertheless, this crucial implication was disregarded.

The most interesting part of the solutions in \mybf{Case I} and \mybf{Case II} is that there exists an exact Schwarzschild solution with non-trivial parameters $\xi$ and $b$ and non-trivial field configurations.
Since $B_\mu$ is a real-valued vector field, we can easily see that $\alpha\geq 1$ in the space-like cases and $\alpha>0$ in the light-like cases, hence the solutions reduce to the Schwarzschild solution only when $\ell\equiv\xi b^2=0$.
However, in the time-like cases, $\alpha$ can be zero if $\beta^2=1$ in \mybf{Case I} and $(\beta+\gamma)^2=1$ in \mybf{Case II}.
We can state this result from another point of view, and it appears to be more intriguing.
As we discussed in the above paragraph, the solutions of bumblebee gravity in this work are also solutions to a model in which a vector field \emph{non-minimally} coupled to Einstein gravity.
It is well-known that in general relativity, the Schwarzschild spacetime is only a solution of vacua, and in the presence of matter \emph{minimally} coupled to gravity, the Schwarzschild metric is not a solution any longer.
On the other hand, if the vector field is minimally coupled to gravity, then its gauge degree of freedom recovers, and we can interpret it as an electromagnetic field, leading to the usual Reissner-Nordstr{\"o}m solution.
In contrast, 
when the vector field is non-minimally coupled to gravity as~\eq{eq:vt}, the Schwarzschild metric is shown to be a solution to the Einstein equation, even if we are not dealing with vacua, {\it i.e.,} there are non-vanishing vector fields distributed through the spacetime.
For example, in \mybf{Case II} and for time-like VEVs, the non-zero components of the field strength are \(F_{tr}=-F_{rt}=-\gamma b R_s/r^2\), with a charge of \(Q=\gamma b R_s\).
In conclusion, the Schwarzschild can also be used to describe a spacetime formed by a vector field with specific field configurations non-minimally coupled to gravity.

For the tests of bumblebee gravity based on the vacuum solution, the only testable variable is the combination $\alpha \ell$.
The difference between the solutions with $b_\mu$, which has two-component VEVs, and the solution in Ref~\cite{Casana:2017jkc} is that the $\ell$ is replaced with $\alpha\ell$.
With the same analysis, the constraint from the solar system is $\alpha\ell\leq 10^{-13}$.
However, since $\alpha$ can be zero in the time-like case and the metric is of the Schwarzschild form,
the experimental tests within the solar system would be invalidated.
% Therefore, for the time-like bumblebee gravity, 
Therefore, testing of bumblebee gravity requires methods beyond those relying on vacuum solutions, such as dynamical approaches including gravitational waves.

\section*{Summary}

\del{
In summary, in this work, we solve the static spherical vacuum problems in bumblebee gravity with the most general forms of VEVs.
Based on the systematic solving procedure, we propose a complete classification of the solutions.
We find new exotic solutions under a special set of parameters. We find that when $\xi=\kappa/2$ and $b^2=2/\kappa$, the solution can be parameterized by a function, and thus the theory is degenerate and ill-defined.
Those exotic solutions are also solutions to the theory of a massless vector field non-minimally coupled to gravity with a general coupling constant $\xi$,
and when $\xi=\kappa/2$, this vector-tensor theory is also degenerate and ill-defined.
The most interesting and important part of the solutions is that there exists an exact Schwarzschild solution with non-trivial parameters $\xi$ and $b$ and non-trivial field configurations in the time-like cases.
The exact Schwarzschild solution poses a challenge to experimental constraints on bumblebee gravity relying on vacuum solutions, such as the tests based on the advance of the perihelion, the bending of light, and the time delay of light.
Therefore, testing of bumblebee gravity requires methods beyond those relying on vacuum solutions.
}

\add{
In summary, in this work, we solve the static spherical vacuum problems in bumblebee gravity with the most general forms of VEVs.
Based on the systematic solving procedure, we propose a complete classification of the solutions.
We find new exotic solutions under a special set of parameters. We find that when $\xi=\kappa/2$ and $b^2=2/\kappa$, the solution can be parameterized by a function, and thus the theory is degenerate and ill-defined.
Those exotic solutions are also solutions to the theory of a massless vector field non-minimally coupled to gravity with a general coupling constant $\xi$,
and when $\xi=\kappa/2$, this vector-tensor theory is also degenerate and ill-defined.
}

\add{
It is also worth emphasizing the case $b^\mu b_\mu=0$, which corresponds to a light-like vacuum expectation value of the bumblebee field.
% while lying outside the spontaneous symmetry breaking regime, may also admit nontrivial static spherical solutions.
This situation represents a qualitatively distinct regime of the model, in which the background vector becomes null while still producing nontrivial modifications to the gravitational field equations.
Within our classification, consistent static spherical solutions with light-like VEVs are explicitly obtained, demonstrating that this previously less-explored sector of the theory admits nontrivial geometries.
}

\add{
The most interesting and important part of the solutions is that there exists an exact Schwarzschild solution with non-trivial parameters $\xi$ and $b$ and non-trivial field configurations in the time-like cases.
The exact Schwarzschild solution poses a challenge to experimental constraints on bumblebee gravity relying on vacuum solutions, such as the tests based on the advance of the perihelion, the bending of light, and the time delay of light.
Therefore, testing of bumblebee gravity requires methods beyond those relying on vacuum solutions.
}

\emph{\underline{Note added.}} 
% During the preparation of this manuscript, we became aware of Ref.~\cite{Liu:2025oho}, which also found similar solutions for general light-like and space-like VEVs.
% They focused on analyzing the thermodynamics of these solutions, while we analyze them from a different point of view.
% We also showed that they have not found all solutions, as addressed in this work.
During the preparation of this manuscript, we became aware of Ref.~\cite{Liu:2025oho}, which also reported similar solutions for general light-like and space-like VEVs.
In particular, Ref.~\cite{Liu:2025oho} obtained the same light-like and space-like solutions as our Case I and Case II, and focused on analyzing their thermodynamic properties.
In contrast, our work explores the full range of possible solutions across different parameter regimes.
Furthermore, we identify an additional space-like case (Case V) and obtain more general time-like solutions, and we discuss their physical implications.

% If there are no assumptions on $\xi$ and $b^2$, 

\section*{Acknowledgements}

This work was supported in part by the National Natural Science Foundation of China under Grant No.~12547101. HL was also supported by the start-up fund of Chongqing University under No.~0233005203009, and JZ was supported by the start-up fund of Chongqing University under No.~0233005203006.

\appendix

\section{Derivation of the Explicit Analytical Solution}\label{appendix}

% Without losing generality, we assume the static spherical-symmetric metric to be
% \begin{equation}
%     ds^2=-e^{2\alpha(r)}dt^2+e^{-2\alpha(r)}dr^2+R(r)^2d\Omega^2, \label{eq:met}
% \end{equation}
% and the general static spherical background $b_\mu$ can be expressed as
% \begin{equation}
%     b_\mu=(b_t,b_r,0,0),\label{eq:b_s}
% \end{equation}
% where $b_t$ and $b_r$ are functions of $r$.
Substituting \eq{eq:met} and \eq{eq:b_s} into \eq{eq:gt} and \eq{eq:b}, 
we can obtain five equations: three from the $tt$, $rr$ and $\theta\theta$ component of the equation of motions for $g_{\mu\nu} $ from \eq{eq:gt}, and two from the $t$ and $r$ component of the equation of motions for $B_\mu$ from \eq{eq:b}.
The $\phi\phi$ component of \eq{eq:gt} does not matter because it is proportional to the $\theta\theta$ component.
Also, the $tr$ component of \eq{eq:gt} is proportional to the $r$ component of \eq{eq:b}.
Here we denote the left-hand sides of the  $tt$, $rr$ and $\theta\theta$ component of \eq{eq:gt} as $\mathcal{E}_{tt}$, $\mathcal{E}_{rr}$ and $\mathcal{E}_{\theta\theta}$, and the left-hand sides of the $t$ and $r$ component of \eq{eq:b} as $\mathcal{E}_{t}$, $\mathcal{E}_{r}$, respectively.
% Here we denote $tt$, $rr$ and $\theta\theta$ component of \eq{eq:gt} as $\mathcal{E}_{tt}=0$, $\mathcal{E}_{rr}=0$ and $\mathcal{E}_{\theta\theta}=0$, and $t$ and $r$ component of \eq{eq:b} as $\mathcal{E}_{t}=0$, $\mathcal{E}_{r}=0$, respectively.
% Due to the complexity and length of the equations, we will only outline the solution method, rather than writing out the specific equation formulas for every step. 

To simplify the expression, we define $\ell\equiv\xi b^2$.

\subsection{Solutions for Space-like VEVs}

% Without losing generality, we assume the static spherical metric to be
% \begin{equation}
%     ds^2=-e^{2\alpha(r)}dt^2+e^{-2\alpha(r)}dr^2+R(r)^2d\Omega^2. \label{eq:met}
% \end{equation}

For space-like VEVs, $b_\mu b^\mu - b^2=0$.
To satisfy this constraint, we can write the components of the general static spherical background $b_\mu$ as
% \begin{equation}
%     b_\mu=(be^{\alpha(r)} \sinh(U(r)),be^{-\alpha(r)}\cosh(U(r)),0,0),\label{eq:b_s}
% \end{equation}
\begin{align}
    % &b_\mu=(b_t,b_r,0,0),\\ \label{eq:b_s}
    &b_t=be^{\alpha(r)} \sinh(U(r)),\\
    &b_r=be^{-\alpha(r)}\cosh(U(r)),
\end{align}
where $U(r)$ is a function to be determined from the equations of motion.
% We now have five equations: three from the $tt$, $rr$ and $\theta\theta$ component of the equation of motions for $g_{\mu\nu} $ \eq{eq:gt}, and two from the $t$ and $r$ component of the equation of motions for $B_\mu$ \eq{eq:b}.
However, we only have three variables: $\alpha(r)$, $R(r)$ and $U(r)$.
Therefore, the problem is overdetermined, and we need to find a solution that satisfies all the equations.
The equations are
\begin{equation}
\begin{aligned}
    \mathcal{E}_{tt}=&R(r) \left[2 \left(4 \ell-b^2 \kappa \right) \alpha ' U' \sinh (2 U)\right.\\
    &-U'^2 \left(b^2 \kappa +\left(b^2 \kappa -4 \ell\right) \cosh (2
   U)\right)\\
   &+\alpha '^2 \left(b^2 \kappa -b^2 \kappa  \cosh (2 U)+4 \ell+8\right)\\
   &+\left.2 (\ell+2) \alpha ''+2 \ell U'' \sinh (2 U)\right]\\
   &-4 \ell R'' \cosh ^2(U)\\
   &+4 R' \left((\ell+2) \alpha '+\ell U' \sinh (2 U)\right),
\end{aligned}
\end{equation}
\begin{equation}
\begin{aligned}
\mathcal{E}_{rr}=& R \left[-2 \left(4  \ell-b^2 \kappa \right) \alpha ' U'\sinh (2 U)\right.\\
&+U'^2 \left(b^2 \kappa +\left(b^2 \kappa -4  \ell\right) \cosh (2U\right)\\
&-\alpha '^2 \left(b^2 \kappa +\left(8  \ell-b^2 \kappa \right) \cosh (2 U)+4  \ell+8\right)\\
&-2 \left( \ell U'' \sinh (2 U)\right.
+\left.\left.\alpha ''(2  \ell \cosh (2 U)+ \ell+2)\right)\right]\\
&-4 R'' \left(3  \ell \cosh ^2(U)+2\right)\\
&-4 R' \left[ \ell U' \sinh (2 U)\right.\\
&+\left.\alpha ' (2  \ell \cosh (2U)+ \ell+2)\right],
\end{aligned}
\end{equation}
\begin{equation}
\begin{aligned}
\mathcal{E}_{\theta\theta}=& R^2 \left[\alpha '^2 \left(b^2 \kappa -b^2 \kappa  \cosh (2 U)+4 \ell\right)\right.\\
&-2 b^2 \kappa  \alpha ' U' \sinh (2 U)\\
&\left.-2 b^2 \kappa  U'^2 \cosh ^2(U)+2 \ell \alpha ''\right]\\
&-4 R R' \left[\ell U' \sinh (2 U)\right.\\
&\left.+\alpha ' (\ell \cosh (2 U)+2)\right]\\
&-2 R'^2 (\ell \cosh (2 U)+\ell+2)\\
&+4 e^{-2 \alpha }-4 R R'',
\end{aligned}
\end{equation}
\begin{equation}
\begin{aligned}
\mathcal{E}_{t}=& 2 R' \left[\left(b^2 \kappa -\ell\right) \alpha ' \sinh (U)\right.\\
&\left.+b^2 \kappa  U' \cosh (U)\right]\\
&+R \left[-\left(2 \ell-b^2 \kappa \right) \alpha '^2 \sinh (U)\right.\\
&+b^2 \kappa  U'' \cosh (U)
+2 b^2 \kappa  \alpha ' U' \cosh (U)\\
&+b^2 \kappa  U'^2 \sinh (U)
+b^2 \kappa  \alpha '' \sinh (U)\\
&\left.-\ell \alpha '' \sinh (U)\right],
\end{aligned}
\end{equation}
\begin{equation}
\begin{aligned}
\mathcal{E}_{r}=& 2 R''+2 \alpha ' R'+R \left(\alpha ''+2 \alpha '^2\right).
\end{aligned}
\end{equation}

% Substituting \eq{eq:met} and \eq{eq:b_s} into \eq{eq:gt} and \eq{eq:b}, we can obtain the equations for $\alpha(r)$, $R(r)$ and $U(r)$.
% Due to the complexity and length of the equations, we will only outline the solution method, rather than writing out the specific equation formulas for every step. 
% % Here we denote $tt$, $rr$ and $\theta\theta$ component of \eq{eq:gt} as $\mathcal{E}_{tt}$, $\mathcal{E}_{rr}$ and $\mathcal{E}_{\theta\theta}$, and $t$ and $r$ component of \eq{eq:b} as $\mathcal{E}_{t}$, $\mathcal{E}_{r}$, respectively.
% To simplify the notation, we define $\ell=\xi b^2$.
Firstly, by a linear combination of $\mathcal{E}_{tt}$, $\mathcal{E}_{rr}$ and $\mathcal{E}_{r}$, we can get
\begin{equation}
    (1+\ell)\cosh(U(r))R''(r)=0. \label{eq:ellRs}
    % \cosh(U(r))R''(r)=0.
\end{equation}
If $\ell\neq -1$, then $R''(r)=0$. 
Without losing generality, the solution is $R(r)=r$, since we can always redefine $t$ and $r$ such that the metric retains the form as \eq{eq:met}.% and $\cosh(U(r)\neq 0$.
With the solution of $R(r)$, $\mathcal{E}_{r}$ becomes
\begin{equation}
    2\alpha'(r)+r(2\alpha'(r)^2+\alpha''(r))=0,
\end{equation}
and the solution is
\begin{equation}
    \alpha(r)=\frac{1}{2}\log\left(\frac{1}{A}\left(1-\frac{R_s}{r}\right)\right),
\end{equation}
where $A$ and $R_s$ is constants.
With the solution of $R(r)$ and $\alpha(r)$, $\mathcal{E}_{t}$ becomes
\begin{align}
    &-\frac{R_s^2\sinh(U)}{4(r-R_s)^2r}+\frac{\cosh(U)(2r-R_s)U'}{r-R_s}\nonumber\\
    &+r\sinh(U)U'^2+r\cosh(U)U''=0.
\end{align}
With the transformation $V = \sinh(U)$, the equation is given by
\begin{equation}
    V''+\frac{2r-R_s}{(r-R_s)r}V'-\frac{R_s^2}{4(r-R_s)^2r^2}V=0,\label{eq:V}
\end{equation}
and the solution is
% \begin{equation}
%     U(r)=\arcsinh\left(\frac{\beta  r+\gamma (r-R_s)}{\sqrt{r(r-R_s)}}\right).
% \end{equation}
\begin{align}
    V&=\frac{\beta  r+\gamma (r-R_s)}{\sqrt{r(r-R_s)}},\\
    U&=\arcsinh(V).
\end{align}
So far, $\mathcal{E}_{t}$ and $\mathcal{E}_{r}$ are satisfied by the current solution, and we need to check the other three equations.
Substituting the solution of $R(r)$, $\alpha(r)$ and $U(r)$ into $\mathcal{E}_{tt}$, $\mathcal{E}_{rr}$ and $\mathcal{E}_{\theta\theta}$, we can find they are satisfied if and only if
\begin{align}
&A=1+\ell (1+(\beta +\gamma )^2), \label{eq:sol_As}\\
&R_s^2(2\ell-\kappa b^2)\gamma ^2=0.\label{eq:cons_s}
\end{align}

% \eq{eq:sol_As} implies that 
% \begin{equation}
%     A=\frac{1}{1+(1+(\beta +\gamma )^2)\ell}.
% \end{equation}
For non-zero $R_s$, if $2\ell-\kappa b^2\neq0$, then \eq{eq:cons_s} implies $\gamma =0$, and the solution for $g_{\mu\nu}$ is
\begin{equation}
\begin{aligned}
    ds^2&=-\frac{1}{A}fdt^2+A\frac{1}{f}dr^2+r^2d\Omega^2,\\ \label{eq:solg}
    % A&=1+\ell (1+(\beta +\gamma )^2),
\end{aligned}
\end{equation}
and for $b_\mu$ is
\begin{equation}
\begin{aligned}
    b_t=&  \frac{b}{\sqrt{A}}\beta ,\\
    % b_r=& b\sqrt{(b_t/b)^2/f^2+1/f},
    b_r=& b\sqrt{A} \frac{\sqrt{f+\beta ^2}}{f},
\end{aligned}  
\end{equation}
where 
%$A=1+(1+\beta ^2)\ell$ and $f=1-R_s/r$.
\begin{equation}
\begin{aligned}
    A&=1+(1+\beta ^2)\ell,\\
    f&=1-\frac{R_s}{r}.
\end{aligned}
\end{equation}
In this solution, $b_t$ is a constant,
and if $\beta =0$, then it becomes the solution obtained in Ref.~\cite{Casana:2017jkc}.
However, for the obtained solution to be physically meaningful, we require $b_r$ to be a real number.
Since $0 < f < 1$ outside the horizon, and $-\infty < f < 0$ inside the horizon,
the solution is always meaningful outside the horizon, but not always meaningful inside the horizon.

% Also, this solution is exactly the solution obtained in Ref.~\cite{Liu:2025oho}, up to a redefinition of $t$.

If $2\ell-\kappa b^2=0$, i.e., $\xi=\kappa/2$, then the solution for the metric is still \eq{eq:solg}, with $A=1+(1+(\beta +\gamma )^2)\ell$,
% \begin{align}
%     ds^2&=-fdt^2+\frac{1}{f}dr^2+r^2d\Omega^2,\\
%     f&=\frac{1}{1+(1+(\beta +\gamma )^2)\ell}\left(1-\frac{R_s}{r}\right),
% \end{align}
and the solution for $b_\mu$ is
\begin{equation}
\begin{aligned}
    b_t=&  \frac{b}{\sqrt{A}}\left(\beta +\gamma f\right),\\
    b_r=& b\sqrt{A} \frac{\sqrt{(\beta +\gamma f)^2+f}}{f}.
\end{aligned}    
\end{equation}
In this case, to let $b_r$ be a real number inside the horizon, we need $\beta \gamma \leq-1/4$.
This solution is similar to the solution above, but $b_t$ is not a constant, and $b_\mu$ has more choices when the background metric is fixed, and the number of LV parameters is three.
In this case, $b_{\mu\nu}$ has non-vanishing component
\begin{equation}
    b_{tr}=-b_{rt}=-\frac{\gamma  b R_s}{\sqrt{A}}\frac{1}{r^2},
\end{equation}
and this component vanishes if $\gamma =0$.

\subsection{Solutions for Light-like VEVs}

For light-like VEVs, $b_\mu b^\mu=0$.
To satisfy this constraint, we can write the components of the general static spherical background $b_\mu$ as
\begin{align}
    % &b_\mu=(b_t,b_r,0,0),\\ \label{eq:b_s}
    &b_t=be^{\alpha(r)} U(r),\\
    &b_r=be^{-\alpha(r)}U(r),
\end{align}
where $U(r)$ is a function to be determined from the equations of motion.
The equations are
\begin{equation}
\begin{aligned}
\mathcal{E}_{tt}=&U^2 \left(b^2 \kappa  R \alpha '^2+2 \ell R''\right)\\
&-2 U\left(R \left(\left(4 \ell-b^2 \kappa \right) \alpha ' U'+\ell U''\right)+2 \ell R' U'\right)\\
&+R \left(\left(b^2 \kappa -2 \ell\right) U'^2-2\left(\alpha ''+2 \alpha '^2\right)\right)\\
&-4 \alpha ' R',
\end{aligned}
\end{equation}
\begin{equation}
\begin{aligned}
\mathcal{E}_{rr}=&2 U \left(R \left(\left(4 \ell-b^2 \kappa \right) \alpha ' U'+\ell U''\right)+2 \ell R' U'\right)\\
&+U^2 \left(R \left(\left(8 \ell-b^2 \kappa \right) \alpha '^2+4 \ell \alpha ''\right)+6 \ell R''+8 \ell \alpha ' R'\right)\\
&-R \left(\left(b^2 \kappa -2 \ell\right) U'^2-2 \left(\alpha ''+2 \alpha '^2\right)\right)\\
&+4 R''+4 \alpha ' R',
\end{aligned}
\end{equation}
\begin{equation}
\begin{aligned}
\mathcal{E}_{\theta\theta}=&-b^2 \kappa  R^2 \left(U'+U \alpha '\right)^2\\
&-4 R R' \left(\ell U U'+\ell U^2 \alpha '+\alpha '\right)\\
&-2 \left(\ell U^2+1\right) R'^2+2 e^{-2 \alpha }-2 R R'',
\end{aligned}
\end{equation}
\begin{equation}
\begin{aligned}
\mathcal{E}_{t}=&2 R' \left(U \left(b^2 \kappa -\ell\right) \alpha '+b^2 \kappa  U'\right)\\
&+R \left[U \left(\left(b^2 \kappa -\ell\right) \alpha ''+\left(b^2 \kappa -2 \ell\right) \alpha '^2\right)\right.\\
&\left.+b^2 \kappa  U''+2 b^2 \kappa  \alpha ' U'\right],
\end{aligned}
\end{equation}
\begin{equation}
\begin{aligned}
\mathcal{E}_{r}=& 2 R''+2 \alpha ' R'+R \left(\alpha ''+2 \alpha '^2\right).
\end{aligned}
\end{equation}
Still, a linear combination of $\mathcal{E}_{tt}$, $\mathcal{E}_{rr}$ and $\mathcal{E}_{r}$ gives 
\begin{equation}
    U(r)R''(r)=0,
\end{equation}
which provides $R(r)=r$ since $U(r)\neq 0$.
And also, $\mathcal{E}_{r}$ becomes
\begin{equation}
    2\alpha'(r)+r(2\alpha'(r)^2+\alpha''(r))=0,
\end{equation}
and the solution is
\begin{equation}
    \alpha(r)=\frac{1}{2}\log\left(\frac{1}{A}\left(1-\frac{R_s}{r}\right)\right),
\end{equation}
where $A$ and $R_s$ is constants.
In this case, $\mathcal{E}_{t}$ becomes
\begin{equation}
    U''+\frac{2r-R_s}{(r-R_s)r}U'-\frac{R_s^2}{4(r-R_s)^2r^2}U=0,
\end{equation}
as the same equation as the space-like case after the transformation $V=\sinh U$,
and the solution is
\begin{equation}
    U(r)=\frac{\beta  r+\gamma (r- R_s)}{\sqrt{r(r-R_s)}}.
\end{equation}
And now to let $\mathcal{E}_{tt}$, $\mathcal{E}_{rr}$ and $\mathcal{E}_{\theta\theta}$ be zero, we have
\begin{align}
1+\ell (\beta+\gamma) ^2&=A, \label{eq:sol_Al}\\
R_s^2(2\ell-\kappa b^2)\gamma ^2&=0.\label{eq:cons_l}
\end{align}
Still, to have a solution, we need $\gamma =0$ or $\xi=\kappa/2$.
The solution is still in the same form as \eq{eq:solg}.
% However, in both cases, the solution to $g_{\mu\nu}$ is the same as \eq{eq:solg}, with $A=1+\beta ^2\ell$.
% \begin{align}
%     ds^2&=-fdt^2+\frac{1}{f}dr^2+r^2d\Omega^2,\\
%     f&=\frac{1}{1+\beta ^2\ell}\left(1-\frac{R_s}{r}\right).
% \end{align}

If $\xi\neq\kappa/2$, then $\gamma$ needs to be zero.
With $A=1+\ell \beta^2$, the solution for $b_\mu$ is
\begin{align}
    b_t=& b \frac{\beta }{\sqrt{A}},\\
    b_r=& b \frac{\beta  \sqrt{A}}{f}.
\end{align}
If $\xi=\kappa/2$, then with $A=1+\ell (\beta+\gamma)^2$, the solution can be extended as
\begin{align}
    b_t=& b \frac{1}{\sqrt{A}}\left(\beta +\gamma  f\right),\\
    % b_r=& \frac{b_t}{f},
    b_r=& b\sqrt{A}\left(\frac{\beta }{f}+\gamma  \right),
\end{align}
and $b_{\mu\nu}$ has non-vanishing component
\begin{equation}
    b_{tr}=-b_{rt}=-\frac{\gamma  b R_s}{\sqrt{A}}\frac{1}{r^2}.
\end{equation}
For both cases, the solution for $b_\mu$ is real outside and inside the horizon.

\subsection{Solutions for Time-like VEVs}\label{sec:timelike}
For time-like VEVs, $b_\mu b^\mu=-b^2$, thus we can write $b_\mu$ as
\begin{align}
    % &b_\mu=(b_t,b_r,0,0),\\ \label{eq:b_s}
    &b_t=be^{\alpha(r)} \cosh(U(r)),\\
    &b_r=be^{-\alpha(r)}\sinh(U(r)).
\end{align}
The equations are
\begin{equation}
\begin{aligned}
\mathcal{E}_{tt}=&-R \left[-2 \left(4 \ell-b^2 \kappa \right) \alpha ' U' \sinh (2 U)\right.\\
&+U'^2 \left(\left(b^2 \kappa -4 \ell\right) \cosh (2 U)-b^2 \kappa \right)\\
&+\alpha '^2 \left(b^2 \kappa +b^2 \kappa  \cosh (2 U)+4 \ell-8\right)\\
&+2 \ell \alpha ''-2 \ell U'' \sinh (2 U)-4 \alpha ''\left.\right]\\
&-4 \ell R'' \sinh ^2(U)\\
&+4 R' \left(\ell U' \sinh (2 U)-(\ell-2) \alpha '\right),
\end{aligned}
\end{equation}
\begin{equation}
\begin{aligned}
\mathcal{E}_{rr}=&R \left[\right.-2 \left(4 \ell-b^2 \kappa \right) \alpha ' U' \sinh (2 U)\\
&+U'^2 \left(\left(b^2 \kappa -4 \ell\right) \cosh (2 U)-b^2 \kappa \right)\\
&+\alpha '^2 \left(b^2 \kappa +\left(b^2 \kappa -8 \ell\right) \cosh (2 U)+4 \ell-8\right)\\
&-2 \left(\ell U'' \sinh (2 U)+\alpha '' (2 \ell \cosh (2 U)-\ell+2)\right)\left.\right]\\
&-4 R'' \left(3 \ell \sinh ^2(U)+2\right)\\
&-4 R' \left(\ell U' \sinh (2 U)+\alpha ' (2 \ell \cosh (2 U)-\ell+2)\right),
\end{aligned}
\end{equation}
\begin{equation}
\begin{aligned}
\mathcal{E}_{\theta\theta}=&R^2 \left[\right.\alpha '^2 \left(b^2 \kappa +b^2 \kappa  \cosh (2 U)+4 \ell\right)\\
&+2 b^2 \kappa  \alpha ' U' \sinh (2 U)+2 b^2 \kappa  U'^2 \sinh ^2(U)+2 \ell \alpha ''\left.\right]\\
&+4 R R' \left(\ell U' \sinh (2 U)+\alpha ' (\ell \cosh (2 U)+2)\right)\\
&+2 R'^2 (\ell \cosh (2 U)-\ell+2)-4 e^{-2 \alpha }+4 R R'',
\end{aligned}
\end{equation}
\begin{equation}
\begin{aligned}
\mathcal{E}_{t}=&2 R' \left[\right.\left(b^2 \kappa -\ell\right) \alpha ' \cosh (U)+b^2 \kappa  U' \sinh (U)\left.\right]\\
&+R \left[\right.-\left(2 \ell-b^2 \kappa \right) \alpha '^2 \cosh (U)\\
&+b^2 \kappa  U'' \sinh (U)+2 b^2 \kappa  \alpha ' U' \sinh (U)\\
&+b^2 \kappa  U'^2 \cosh (U)+b^2 \kappa  \alpha '' \cosh (U)\\
&-\ell \alpha '' \cosh (U)\left.\right],
\end{aligned}
\end{equation}
\begin{equation}
\begin{aligned}
\mathcal{E}_{r}=& 2 R''+2 \alpha ' R'+R \left(\alpha ''+2 \alpha '^2\right).
\end{aligned}
\end{equation}
Still, a linear combination of $\mathcal{E}_{tt}$, $\mathcal{E}_{rr}$ and $\mathcal{E}_{r}$ gives 
\begin{equation}
    (\ell-1)\sinh(U(r))R''(r)=0. \label{eq:ellRt}
    % \sinh(U(r))R''(r)=0.
\end{equation}
However, the solution process becomes different in this case, because $U(r)=0$ is also a solution.
In this case, $b_r=0$, which has been discussed in detail in our previous work~\cite{Li:2025bzo}.

If $\ell\neq1$, in the case of non-trivial $b_r$, we have $R''(r)=0$, and $R(r)=r$.
Once more, $\mathcal{E}_{r}$ becomes
\begin{equation}
    2\alpha'(r)+r(2\alpha'(r)^2+\alpha''(r))=0,
\end{equation}
and the solution is
\begin{equation}
    \alpha(r)=\frac{1}{2}\log\left(\frac{1}{A}\left(1-\frac{R_s}{r}\right)\right),
\end{equation}
where $A$ and $R_s$ is constants.
With the solution of $\alpha$ and $R$, $\mathcal{E}_{t}$ becomes
\begin{align}
    &-\frac{R_s^2\cosh(U)}{4(r-R_s)^2r}+\frac{\sinh(U)(2r-R_s)U'}{r-R_s}\nonumber\\
    &+r\cosh(U)U'^2+r\sinh(U)U''=0.
\end{align}
With the similar transformation $V = \cosh(U)$, the equation the same as \eq{eq:V},
and the solution is
\begin{equation}
    U(r)=\arccosh\left(\frac{\beta  r+\gamma (r-R_s)}{\sqrt{r(r-R_s)}}\right).
\end{equation}
Still, to satisfy all of the equations, we have
\begin{align}
    1+\ell (-1+(\beta +\gamma )^2)&=A, \label{eq:sol_At}\\
    R_s^2(2\ell-\kappa b^2)\gamma ^2&=0.\label{eq:cons_t}
\end{align}

The solution for the metric is still \eq{eq:solg}, but with a different value of $A$.
If $\xi\neq\kappa/2$, then $\gamma =0$, 
% the solution for $g_{\mu\nu}$ is
% \begin{align}
%     ds^2&=-fdt^2+\frac{1}{f}dr^2+r^2d\Omega^2,\\
%     f&=\frac{1}{1+(\beta ^2-1)\ell}\left(1-\frac{R_s}{r}\right),
% \end{align}
with $A=1+(\beta ^2-1)\ell$,
the solution for $b_\mu$ is
\begin{equation}
\begin{aligned}
    b_t=&  \frac{b}{\sqrt{A}}\beta ,\\
    b_r=& b\sqrt{A} \frac{\sqrt{\beta ^2-f}}{f}.
 \end{aligned}
 \end{equation}
 
 In this case, all components of $b_{\mu\nu}$ are zero.
For the solution to be physically meaningful inside the horizon, we need $\beta ^2\geq 1$.
 Surprisingly, when $\beta =1$, then $A=1$, and the solution for the metric reduces to the Schwarzschild solution. The non-trivial solution for $b_\mu$ is
 \begin{equation}
% b_\mu = (b, b\sqrt{\frac{R_s}{r}}\left(1-\frac{R_s}{r}\right)^{-1},0,0).
b_\mu = (b, b\frac{\sqrt{R_s r}}{r-R_s},0,0).
 \end{equation}

If $\xi=\kappa/2$, 
% the solution for the metric is
% \begin{equation}
% \begin{aligned}
%     ds^2&=-fdt^2+\frac{1}{f}dr^2+r^2d\Omega^2,\\
%     f&=\frac{1}{1+((\beta +\gamma )^2-1)\ell}\left(1-\frac{R_s}{r}\right),
% \end{aligned}
% \end{equation}
with $A=1+((\beta +\gamma )^2-1)\ell$, 
the solution for $b_{\mu}$ is
\begin{equation}
\begin{aligned}
    b_t=&  \frac{b}{\sqrt{A}}\left(\beta +\gamma f\right),\\
    b_r=& b\sqrt{A} \frac{\sqrt{(\beta +\gamma f)^2-f}}{f}.
\end{aligned}
\end{equation}
In this case, $b_{\mu\nu}$ has non-vanishing component
\begin{equation}
    b_{tr}=-b_{rt}=-\frac{\gamma  b R_s}{\sqrt{A}}\frac{1}{r^2}.
\end{equation}
For the solution to be physically meaningful inside the horizon, we need $(\beta +\gamma )^2\geq 1$, together with $\beta  \gamma \geq1/4$ or $\beta \gamma +\gamma ^2\leq 1/2$.
Also, when $\beta +\gamma =1$, this solution reduces to the Schwarzschild solution, and in this case, the non-trivial $b_\mu$ is
\begin{equation}
\begin{aligned}
b_t&=b\left(1-\frac{\gamma R_s}{r}\right),\\
b_r&=b\frac{\sqrt{R_s(\gamma ^2 R_s +(1-2\gamma )r)}}{r-R_s}.
\end{aligned}
\end{equation}
In order for $b_r$ to be a real number, we need $\gamma \leq 1/2$.

\section{Series solutions with \(\ell=\pm 1\)}\label{sec:series}

From \eq{eq:ellRs} and \eq{eq:ellRt}, we can see that if $\ell=-1$ in the space-like case or $\ell=1$ in the time like case,
$\mathcal{E}_{tt}$, $\mathcal{E}_{rr}$ and $\mathcal{E}_{r}$ are linearly dependent, thus we lose the constraint on $R(r)$, and we may overlook solutions.
However, in these cases, these equations cannot be simplified further.
Motivated by the technical approach of Ref~\cite{Lu:2015cqa}, here we solve the equations using the Taylor expansion method.
The result suggests that we indeed overlook solutions.

% Here, we mainly focus on the time-like case with $\ell=1$,
% and the space-like case with $\ell=-1$ is similar.
Since $\ell=\xi b^2$, in both cases, $ s \xi b^2=1$, where $b_\mu b^\mu+sb^2=0$.
We use a different coordinate system to solve the problem, with the metric to be
\begin{equation}
    ds^2=-G(r)dt^2+\frac{1}{H(r)}dr^2+r^2d\Omega^2.\label{eq:metGH}
\end{equation}
Supposing that there exists a black-hole horizon at some radius $r=R_s>0$, at which the metric functions $G$ and $F$ vanish, we can write near-horizon Taylor expansions of $G$ and $F$ as
\begin{equation}
\begin{aligned}
G(r)&=\frac{1}{R_s}(r-R_s)+\sum_{i=2}^\infty \frac{g_i}{R_s^i}(r-R_s)^i,\\
H(r)&=\frac{h_1}{R_s}(r-R_s)+\sum_{i=2}^\infty \frac{h_i}{R_s^i}(r-R_s)^i,\\ \label{eq:GHseries}
\end{aligned}
\end{equation}
where $g_i$ and $h_i$ are coefficients and $g_1$ is set to be $1$ by the redefinition of $t$.
We also write the Taylor expansions of $b_t$ as
\begin{equation}
b_t(r)= \sum_{i=0}^\infty \frac{b_i}{R_s^i}(r-R_s)^i.\label{eq:bseries}
\end{equation}
From $b_\mu b^\mu=-sb^2$, we can get 
\begin{equation}
    b_r(r)=\sqrt{\frac{b_t(r)^2-sb^2 G(r)}{G(r)H(r)}},
\end{equation}
and use \eq{eq:GHseries} and \eq{eq:bseries}, we can get the Taylor expansions for $b_r$, expressed by $g_i$, $h_i$ and $b_i$. 
In terms of the metric \eq{eq:metGH}, the equation of $\mathcal{E}_{r}$ is
\begin{equation}
    b_r(r) \left(G \left(r G'+4 G\right) H'-r H \left(G'^2-2 G G''\right)\right)=0. \label{eq:GHt}
\end{equation}
With $b_r(r)\neq 0$, from \eq{eq:GHt} we can express the coefficients $h_i ~(i\geq 2)$ as functions of $g_i~(i\geq 2)$ and $h_1$.
Substituting the Taylor expansions of $G(r)$, $H(r)$, $b_t(r)$, $b_r(r)$, together with the relation between $h_i$ and $g_i$, into other equations of motion, we can solve the coefficients.
The results show that any other coefficients can be expressed by $b_0$ and $b_1$, and together with the parameter $R_s$, they provide a three-parameter family of solutions.

With $u=s\kappa b^2$, $k=b_1/b_0$, the first few coefficients for $h_i$ are
\begin{equation}
\begin{aligned}
h_1=&\frac{2u}{(2 + 4 k + k^2 u) \kappa b_0^2},\\
h_2=&\left(\frac{3 k^2 u}{4}-1\right) h_1,\\
h_3=&\left(-\frac{1}{8} k^4 u^2-\frac{1}{12} (20 k+27) k^2 u+1\right)h_1,\\
h_4=&\left(\frac{3 k^6 u^3}{64}+\frac{1}{32} (24 k+13) k^4 u^2\right.\\
&\left.+\frac{1}{6} \left(21 k^2+41
   k+27\right) k^2 u-1\right)h_1,\\
\end{aligned}
\end{equation}
% \begin{equation}
% \begin{aligned}
% h_1=&\frac{2}{(2 + 4 k + k^2 u) b_0^2},\\
% h_2/h_1=&\frac{3 k^2 u}{4}-1,\\
% h_3/h_1=&-\frac{1}{8} k^4 u^2-\frac{1}{12} (20 k+27) k^2 u+1,\\
% h_4/h_1=&\frac{3 k^6 u^3}{64}+\frac{1}{32} (24 k+13) k^4 u^2\\
% &+\frac{1}{6} \left(21 k^2+41
%    k+27\right) k^2 u-1,\\
% \end{aligned}
% \end{equation}
for $g_i$ are
\begin{equation}
\begin{aligned}
g_2=&-1 - \frac{k^2 u}{4},\\
g_3=&\frac{k^4 u^2}{8}+\frac{1}{12} (4 k+3) k^2 u+1,\\
g_4=&-\frac{5}{64} k^6 u^3-\frac{1}{96} (40 k+21) k^4 u^2\\
&-\frac{1}{2} (k+1) k^3 u-1,
\end{aligned}
\end{equation}
and for $b_i$ are
% \begin{equation}
% \begin{aligned}
% b_2=&\left(-\frac{k^3 u}{4}-\frac{1}{2} (k+2) k\right )b_0,\\
% b_3=&\left(\frac{k^5 u^2}{8}+\frac{7}{12} (k+1) k^3 u\right.\\
% &\left.+\frac{1}{2} \left(k^2+2 k+2\right) k\right )b_0,\\
% b_4=&\left(-\frac{5}{64} k^7 u^3-\frac{5}{96} (11 k+9) k^5 u^2\right.\\
% &-\frac{1}{24} \left(29 k^2+48
%    k+24\right) k^3 u\\
%    &\left.-\frac{1}{8} \left(5 k^3+12 k^2+12 k+8\right) k\right )b_0.
% \end{aligned}
% \end{equation}
\begin{equation}
\begin{aligned}
b_2/b_0=&-\frac{k^3 u}{4}-\frac{1}{2} (k+2) k,\\
b_3/b_0=&\frac{k^5 u^2}{8}+\frac{7}{12} (k+1) k^3 u\\
&+\frac{1}{2} \left(k^2+2 k+2\right) k,\\
b_4/b_0=&-\frac{5}{64} k^7 u^3-\frac{5}{96} (11 k+9) k^5 u^2\\
&-\frac{1}{24} \left(29 k^2+48
   k+24\right) k^3 u\\
   &-\frac{1}{8} \left(5 k^3+12 k^2+12 k+8\right) k.
\end{aligned}
\end{equation}
The complicated coefficients suggest that perhaps the solution does not possess a closed-form expression.
The solution is physically meaningful when $h_1>0$.
In the time-like case, $s=1$, and when $b_1=0$, the above solution reduces to the solution in Appendix~\ref{sec:timelike}.
However, in the space-like case, $s=-1$ and $u<0$, and the limit $b_1 \to 0$ does not yield a physically meaningful solution.

% The result of the space-like case with $\ell=1$ is similar.
% The difference is that, with the exception of
% \begin{equation}
%     h_1=\frac{2}{(-2 - 4 k + k^2 u) b_0^2},
% \end{equation}
% the remaining coefficients $h_i/h_1$, $g_i$, and $b_i$ are obtained by the substitution $u \to -u$ (i.e., $b^2\to - b^2$) of the coefficients in the expression.
% The solution is physically meaningful when $h_1>0$, indicating that the limit $b_1 \to 0$ does not yield a physically meaningful solution.

\section{Solutions with \(\ell=\pm 1\) and \(\xi=\kappa/2\)}\label{sec:any}

Here, we explore whether other solution branches have been missed in this scenario.
We use the metric \eq{eq:metGH}, and the configuration for $b_\mu$ is
\begin{equation}
\begin{aligned}
b_t=&b\sqrt{G(r)}V(r),\\
b_r=&b\frac{\sqrt{V(r)^2-s}}{\sqrt{H(r)}},
\end{aligned}
\end{equation}
where $s=1$ denotes the time-like case, while $s=-1$ corresponds to the space-like case.%, $b_\mu b^\mu +sb^2=0$.
The equation for $\mathcal{E}_{r}$ is \eq{eq:GHt}, suggesting that
\begin{equation}
H(r)=c_1 \exp\left(
\int \frac{r \left(G'^2-2 G G''\right)}{G \left(r G'+4 G\right)}dr.\label{eq:solH}
\right )
\end{equation}
With the above relation between $H(r)$ and $G(r)$, the equation for $\mathcal{E}_{\theta\theta}$ becomes
\begin{equation}
    4 r  G^2-s r H \left(r V G'+2 G \left(r V'+V\right)\right)^2=0. \label{eq:sGH}
\end{equation}
Outside the horizon, we have $G(r)>0$ and $H(r)>0$.
So when $s=-1$, the left side of \eq{eq:sGH} is always greater than zero, which means that there is no solution for the space-like case.

When $s=1$, the relations of the parameters of this time-like case are $\xi=\kappa/2$ and $b^2=2/\kappa$.
The solution for \eq{eq:sGH} is
\begin{equation}
    V(r)=\frac{\pm 1}{r\sqrt{G(r)}}\left(c_2+\int \sqrt{\frac{G(r)}{H(r)}}dr\right).\label{eq:solV}
\end{equation}
Surprisingly, substituting \eq{eq:solH} and \eq{eq:solV} into $\mathcal{E}_{tt}=0$, $\mathcal{E}_{rr}=0$  and $\mathcal{E}_{t}=0$,
we can find that these equations are satisfied identically!
In this case, given any function $G(r)$, we can construct a solution with \eq{eq:solH} and \eq{eq:solV}.

For this specific set of model parameters, the vacuum field equations exhibit an uncountable degeneracy in their solution space.
Specifically, the metric can be constructed using an arbitrary function $G(r)$, resulting in a continuum of solutions parameterized by this function.
Such a scenario implies that the dynamical equations are degenerate and lose their ability to constrain the metric tensor.
From a physical perspective, the existence of solutions characterized by an arbitrary function is pathological, as it renders the initial value problem ill-posed and the theory non-predictive. Consequently, this specific set of parameters must be excluded from the physically viable parameter space, representing a singular boundary where the underlying theory lacks a well-defined dynamical structure.

\bibliography{refs}

\end{document}